\documentclass[prb,amsmath,floats,floatfix,amssymb,superscriptaddress,twocolumn,reprint]{revtex4}
\usepackage{graphicx}
\usepackage{gensymb}
\usepackage{amsmath}
\usepackage{upgreek}
\begin{document}

\title{Dispersive readout of a silicon quantum dot with an accumulation-mode gate sensor}


\author{A. Rossi}
\email[Electronic mail: ]{ar446@cam.ac.uk}
\affiliation{Cavendish Laboratory, University of Cambridge, J.J. Thomson Avenue, Cambridge, CB3 0HE, U.K.}
\author{R. Zhao}
\author{A. S. Dzurak}
\affiliation{School of Electrical Engineering \& Telecommunications, The University of New South Wales, Sydney 2052, Australia}
\author{M. F. Gonzalez-Zalba}
\affiliation{Hitachi Cambridge Laboratory, J.J. Thomson Avenue, Cambridge, CB3 0HE, U.K.}
\date{\today}

\begin{abstract}
Sensitive charge detection has enabled qubit readout in solid-state systems. Recently, an alternative to the well-established charge detection via on-chip electrometers has emerged, based on in situ gate detectors and radio-frequency dispersive readout techniques. This approach promises to facilitate scalability by removing the need for additional device components devoted to sensing. Here, we perform gate-based dispersive readout of an accumulation-mode silicon quantum dot. We observe that the response of an accumulation-mode gate detector is significantly affected by its bias voltage, particularly if this exceeds the threshold for electron accumulation. We discuss and explain these results in light of the competing capacitive contributions to the dispersive response.
\end{abstract}


\keywords{quantum dots, silicon, gate-based readout, quantum computing, reflectometry.}

\maketitle

\begin{figure}[]
\includegraphics[scale=0.55]{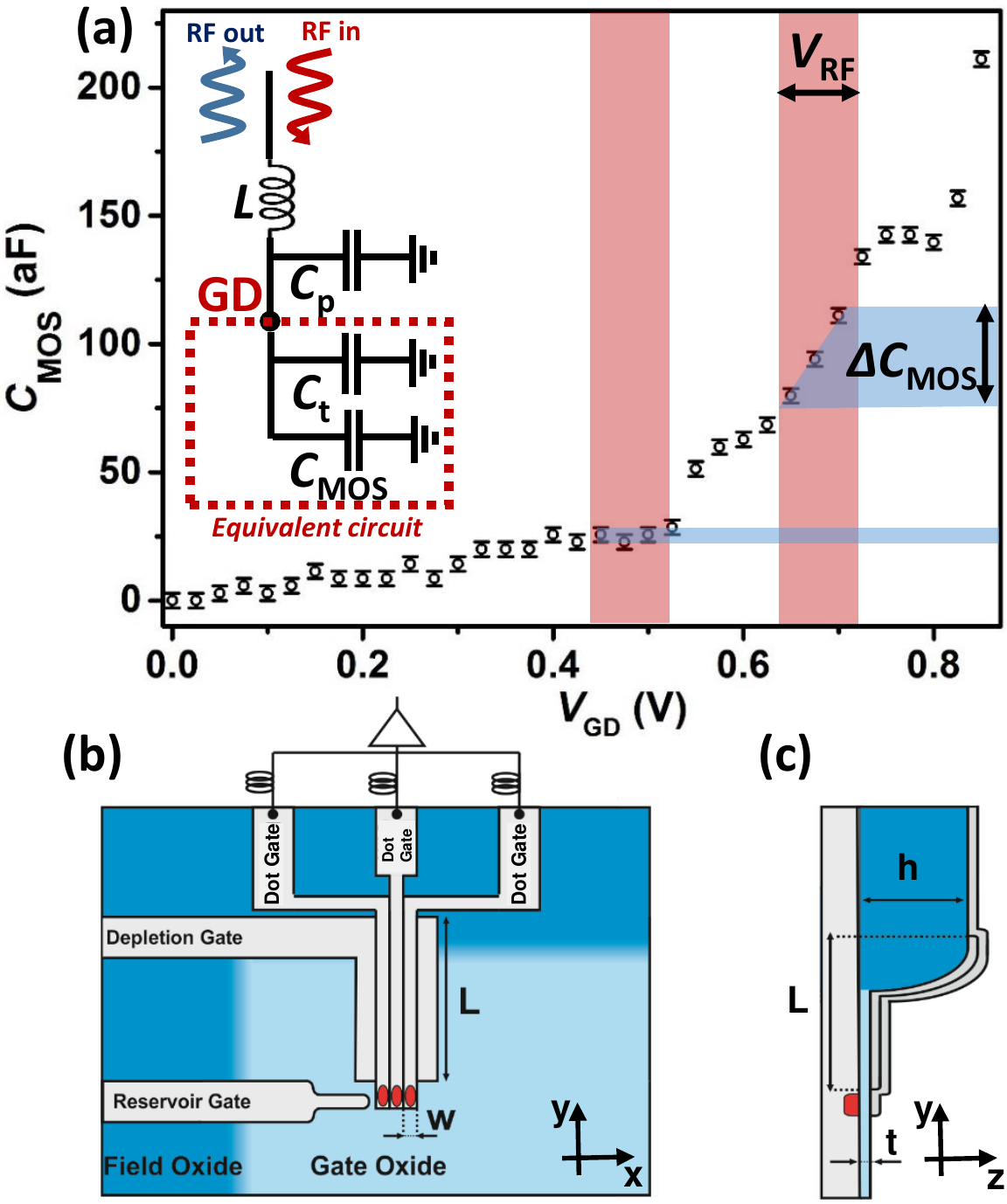}
\caption{(a) MOS capacitance as a function of $V_\textup{GD}$. Error bars represent the measurement resolution. Shaded areas (not to scale for clarity) indicate the peak-to-peak rf amplitude (red) and the relevant  $\Delta C_\textup{MOS}$ variation (blue). Inset: equivalent circuit of the gate detector in series with the resonator circuit. (b) Top view of a device schematic with a linear array of accumulation-mode quantum dots and gate-based readout. The depletion gate runs under the dot gates, of width $w$, for a length $L$. (c) Cross-sectional view of the structure in (b).}
\label{model}
\end{figure}

Reliable measurements of the charge state of nanoscale electronic devices may represent a key ingredient for the realization of future quantum technologies. Typically, non-invasive and sensitive charge readout is achieved by means of on-chip electrometers~\cite{hanson_rev,floris}. The scope of applicability of these sensing techniques is quite broad, ranging from charge noise characterization~\cite{gusta2,mc,ihn} to cryogenic thermometry~\cite{myapl2,mara,mava}, as well as Maxwell's demon implementations~\cite{fuji_dem,pek} and quantum metrology~\cite{kell,lukas,tuomo}. Arguably, one of the research fields that has more largely benefitted from advancements in charge sensing is solid-state quantum information processing~\cite{koppe,petta,pla12,pla13,menno}.\\\indent
\begin{figure}[hb]
\includegraphics[scale=0.425]{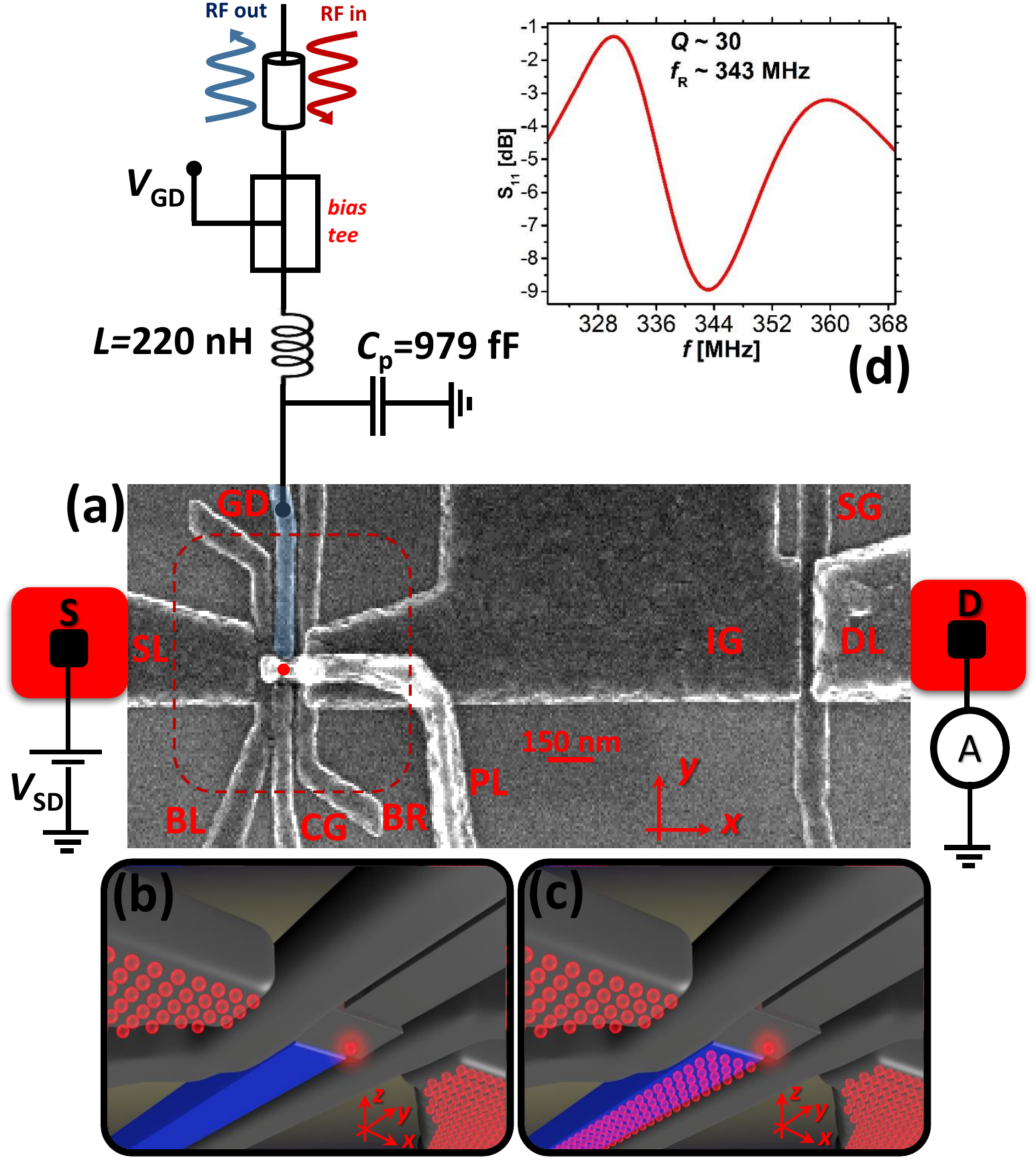}
\caption{(a) SEM image of a device similar to the one used in the experiments and schematic view of the measurement set-up. The gate electrode embedded into the resonant circuit is highlighted in blue. The approximate region where the quantum dot is formed is highlighted with a solid red circle. (b) Artistic illustration of the portion of the device enclosed by the dashed line in (a). Electrons are schematically depicted as red spheres. An isolated electron is representative of the quantum dot position. The gate sensor is shown to operate below threshold voltage. (c) Similar illustration as in (b) except for the gate detector being shown to work above threshold voltage. (d) Characteristic frequency response of the resonator used in the experiments.}
\label{set-up}
\end{figure} 
\begin{figure*}[]
\includegraphics[scale=0.62]{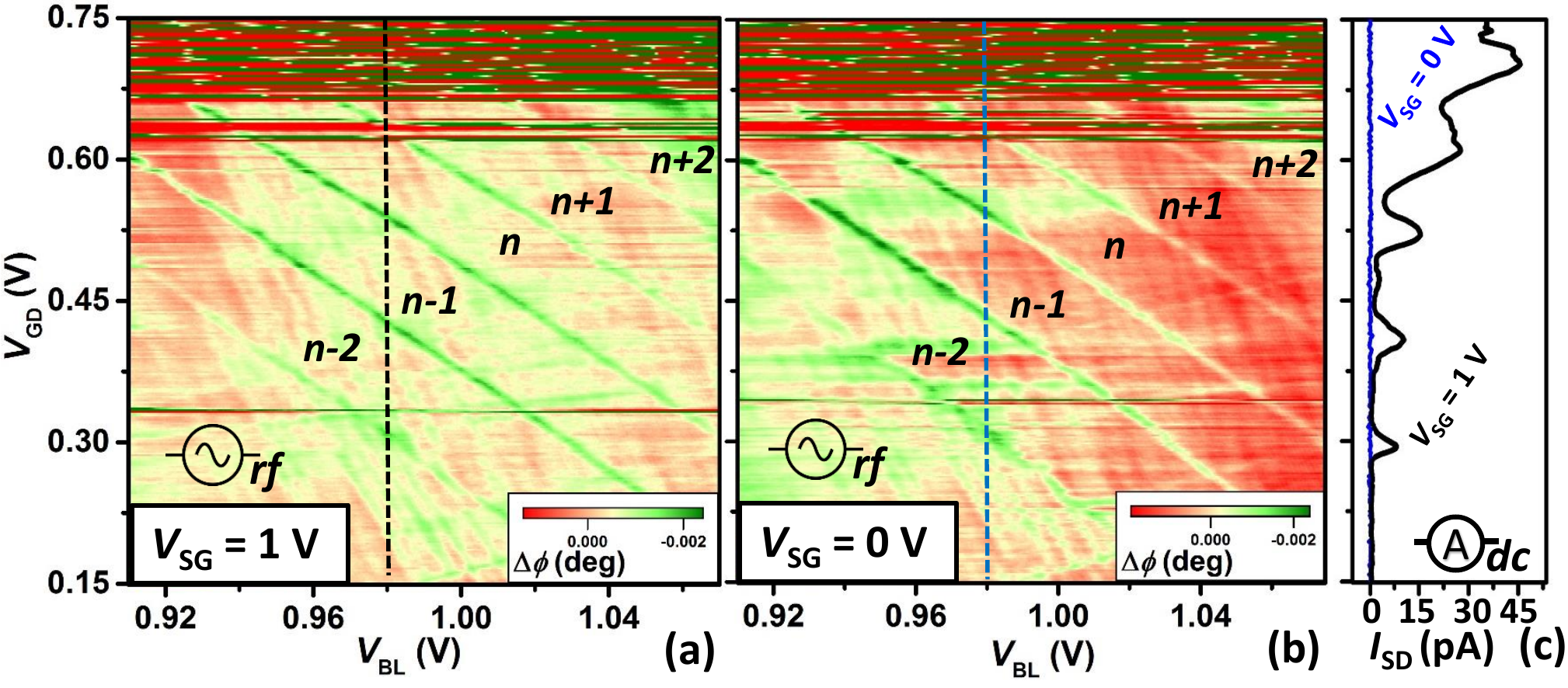}
\caption{(a) Phase response as a function of $V_\textup{BL}$ and $V_\textup{GD}$. Labels indicate charge occupancy of the QD. $V_\textup{SG}=1$~V, $V_\textup{BR}=1.4$~V, $V_\textup{SL}=1$~V, $V_\textup{DL}=V_\textup{IG}=1.4$~V, $V_\textup{SD}=0.4$~mV. (b) Similar plot as in (a) except for $V_\textup{SG}=0$~V. (c) Measurement of the dc current between Source and Drain as a function of $V_\textup{GD}$ at $V_\textup{SG}=0$~V (blue trace) and $V_\textup{SG}=1$~V (black trace). $V_\textup{SD}=0.6$~mV, $V_\textup{BL}=0.98$~V in both cases, as highlighted by the colour-coded dashed lines in panels (a) and (b). All other experimental parameters are unchanged.}
\label{cross}
\end{figure*}
Recently, an alternative approach to implement charge readout has emerged based on radio-frequency (rf) resonant circuit techniques~\cite{schoelko,karl,frey_PRL}. Using in-situ gate electrodes embedded into LC resonators, fast and sensitive charge detection has been attained~\cite{reilly2013,mfgz_nat,mfgz_nano,arjan} by measuring the dispersive shift of the resonator frequency when electron tunneling occurs.  In addition to providing much higher bandwidth than standard electrometry, gate-based reflectometry has an enormous potential for realizing scalable quantum architectures.
In fact, a number of proposals has been already put forward to exploit high frequency techniques to different extents~\cite{line,menno_arx,lieven_arx,reilly2015,child,bur}.\\\indent
Lately, spin-based qubits have been realised in planar silicon-based accumulation-mode quantum dots~\cite{eng,delft_kawa,menno}. Gate-based dispersive readout may be an appealing technique to scale these systems up. In fact, CMOS-compatible architectures have been recently proposed~\cite{menno_arx,lieven_arx} which identify routes toward large scalability of silicon-based spin qubit systems. The suggested readout protocol crucially relies on the gate-based dispersive technique discussed here. It has, therefore, become topical to understand its operational boundaries or limitations. In the specific case of accumulation-mode devices, it has been proposed that the same gate(s) used to define the qubit(s) may be used for dispersive readout~\cite{line,menno_arx}. This would require operating the gate(s) above threshold voltage. However, one has to be aware that the gate capacitance may significantly vary upon bias voltage and, in some circumstances, its contribution may dominate over the tunneling capacitance~\cite{mfgz_arx}. This is a potentially detrimental situation for readout performances. We note that this issue has not been raised before, since previous accounts of dispersive readout were based on either depletion-mode devices~\cite{reilly2013} or etched nanowires~\cite{mfgz_nat, mfgz_nano,arjan}.\\\indent
Here, we perform gate-based dispersive readout of an accumulation-mode quantum dot (QD) fabricated on a planar silicon substrate. By adjusting the dc voltage applied to the gate detector, we observe a significant degradation in the phase response when the bias voltage exceeds the threshold for electron accumulation. We discuss these results in the context of a simple metal-oxide-semiconductor (MOS) capacitance model to explain the voltage-dependent capacitive contributions to the resonator response.
The sample used for this study is a MOS field-effect transistor fabricated on a near-intrinsic natural silicon substrate. Three layers of Al/Al$_y$O$_x$ gates are patterned with electron-beam lithography and deposited on a 8-nm-thick SiO$_2$ gate oxide~\cite{angus,jove}. A scanning electron micrograph (SEM) image of the metal gate stack of a device similar to the one used in the experiments is shown in Fig.~\ref{set-up}(a). Upon application of dc voltages to individual gate electrodes one can locally accumulate a two-dimensional electron gas (2DEG) or form tunnel barriers at the Si/SiO$_2$ interface. We tune these voltages to create a QD approximately in the region highlighted in red in Fig.~\ref{set-up}(a) and represented by an isolated electron in Fig.~\ref{set-up}(b) and (c). This is achieved by forming tunnel barriers with gates BL and BR, and electron reservoirs with gates SL, DL and IG. Note that SL and DL extend to heavily n-type doped regions acting as source (S) and drain (D) ohmics. Gate SG can be used to locally pinch off the drain reservoir when operated below threshold voltage, as we shall discuss later. The gate detector (GD) is highlighted in blue in Fig.~\ref{set-up}(a). It can be dc biased in a similar fashion as the other gates, and, therefore, it allows one to perform charge readout with or without a 2DEG directly formed underneath it. Figures~\ref{set-up}(b) and (c) schematically represent these two operation modes. Gates CG and PL are kept at fixed potentials of 0.22~V and 0~V, respectively.\\\indent 
\begin{figure*}[]
\includegraphics[scale=0.5]{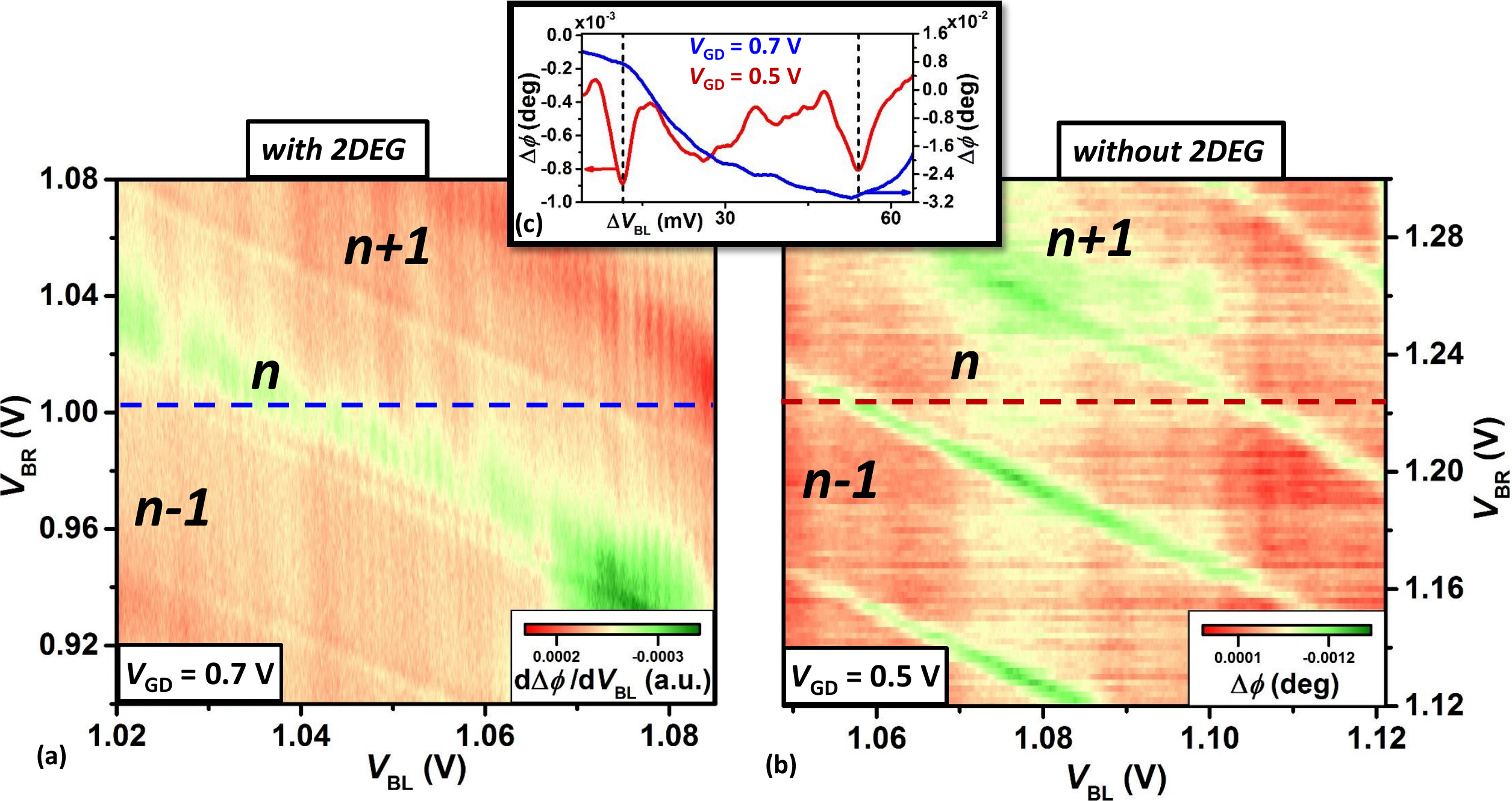}
\caption{(a) Derivative of the phase response as a function of $V_\textup{BL}$ and $V_\textup{BR}$. Labels indicate charge occupancy of the QD. $V_\textup{GD}=0.7$~V, $V_\textup{SG}=0$~V, $V_\textup{SL}=1$~V, $V_\textup{DL}=V_\textup{IG}=1.4$~V, $V_\textup{SD}=0.4$~mV. (b) Phase response as a function of $V_\textup{BL}$ and $V_\textup{BR}$. The other experimental parameters are the same as in (a) except for $V_\textup{GD}=0.5$~V. (c) Phase response as a function of incremental variation of $V_\textup{BL}$ at $V_\textup{GD}=0.7$~V (blue trace, right axis) and $V_\textup{GD}=0.5$~V (red trace, left axis). The red trace is taken from the plot in (b) at $V_\textup{BR}=1.225$~V, the blue trace is taken from the plot in (a) at $V_\textup{BR}=1.005$~V (before derivative is taken). $\Delta V_\textup{BL}=0$ corresponds to $V_\textup{BL}=1.02$~V ($V_\textup{BL}=1.05$~V) for $V_\textup{GD}=0.7$~V ($V_\textup{GD}=0.5$~V). Vertical black dashed lines are guides for the eye to indicate the positions of Coulomb peaks.}
\label{cmp}
\end{figure*}
In order to carry out gate-based dispersive readout, gate GD is embedded in a resonant LC circuit, which is made up of a surface mount inductor ($L=220$~nH) and the device's parasitic capacitance to ground ($C_\textup{p}=979$~fF). The resonator response  (resonant frequency $f_\textup{R}=343$~MHz, quality factor $Q=30$) is shown in Fig.~\ref{set-up}(d). The resonator is connected to a low temperature bias tee [see Fig.~\ref{set-up}(a)] which makes it possible to superimpose a rf signal and a dc offset. Reflectometry is performed at $340$~MHz via homodyne detection of the reflected signal after two stages of amplifications (low temperature and room temperature stages). The phase response, $\Delta\phi$, relates to modifications in the system total capacitance\cite{mfgz_arx}, $\Delta C$, upon variation of the experimental parameters, and it reads $\Delta\phi \approx-\pi Q\Delta C/C_\textup{p}$. This leads to dispersive detection of charge transitions in the dot whenever electron tunnelling occurs. In fact, tunnelling events generate an additional capacitance contribution, known as tunnelling capacitance~\cite{reilly2013,mfgz_nat}, which reads $\Delta C_\textup{t} =\alpha\frac{\partial<ne>}{\partial V_\textup{GD}}$, where $\alpha$ is the gate sensor's lever arm~\cite{SCT} ($\approx$~0.1 eV/V for the device studied here) and $<ne>$ is the average charge of the quantum dot. Therefore, the larger the sensor's lever arm the better it performs in terms of readout sensitivity, which led to a sensitivity as high as 37~$\mu e$/Hz$^{1/2}$ in FinFET transistors~\cite{mfgz_nat}. This was mainly due to an $\alpha$-factor as large as 0.9~eV/V, obtained by using high-k gate dielectrics as thin as 1.3~nm. Besides rf dispersive readout, we also detected the quantum dot charge state via standard dc current measurements. In order to do this, we apply a dc voltage bias ($V_\textup{SD}$) to the source and record the device current at the drain with a digital multimeter after transimpedance amplification at room temperature. All the experiments are performed in a cryogen-free dilution refrigerator at a base temperature of nearly 45 mK.\\\indent 
The dispersive phase signal from the resonator is shown in Fig.~\ref{cross}(a) as a function of $V_\textup{BL}$ and $V_\textup{GD}$. Parallel diagonal lines result from enhanced capacitive contributions at the degenerate charge configurations of the QD. At these points, electron tunneling between the quantum dot and the electron reservoirs is cyclically driven by the oscillatory voltage applied to the resonator. Between each pair of consecutive lines, tunnelling is forbidden due to Coulomb Blockade, and, consequently, the phase response is suppressed. Interestingly, in the upper part of the plot (for $V_\textup{GD}>0.61$~V) the phase signal shows large fluctuations that overshadow the resonant peaks. This region of poor rf readout corresponds to a deterioration in the aspect ratio of the Coulomb peaks, as we observe with the standard dc current measurements shown in Fig.~\ref{cross}(c). The smearing of the peaks and the appearance of a current offset are to be attributed to either the formation of a 2DEG underneath the sensor gate or to enhanced transparency of the QD tunnel barriers. Indeed, the former would provide an additional pathway for the electrons to bypass the dot blockade, while the latter would make co-tunnelling increasingly likely due to loss of quantum confinement.\\\indent
To investigate whether a dc source-drain current, as a mechanism which competes with the rf-driven dot-reservoir electron tunnelling, could be the origin of the large signal background, we switch the current off by setting $V_\textup{SG}$  below threshold. The data are shown in Fig.~\ref{cross}(b). Here, we repeat the same measurement as in panel (a) except for $V_\textup{SG}=0$~V as opposed to $V_\textup{SG}=1$~V. By keeping $V_\textup{SG}$ below threshold, current is prevented from flowing between source and drain due to a discontinuity in the 2DEG which leads to the drain ohmic. This is confirmed by the blue trace in Fig.~\ref{cross}(c). The data in panels (a) and (b) reveal a nearly identical response. This suggests that a dc electric current is not the cause of the readout degradation. 
\\\indent
We now turn to investigate more in detail the region of limited readout in the single-lead configuration. To this end, we set $V_\textup{SG}=0$~V and $V_\textup{GD}=0.7$~V, and measure the phase response of the resonator as a function of $V_\textup{BL}$ and $V_\textup{BR}$. We take the derivative of the raw data to reveal the extremely faint diagonal lines which indicate charge transitions in the QD [see Fig.~\ref{cmp} (a)]. These features would not be well resolved in the bare phase signal because they are buried into a large background which changes as a function of the experimental parameters, as shown in Fig.~\ref{cmp}(c). We point out that, despite the poor contrast, this measurement at relatively high $V_\textup{GD}$ indicates the presence of a properly formed QD. This is confirmed by the charging plot of Fig.~\ref{cmp}(b). In this case, we set $V_\textup{GD}=0.5$~V (within the region of good readout visibility) and measure the phase response at the same operation point as in panel (a). By comparing the two stability diagrams, we observe similar slopes and separations for the charge transitions, and, therefore, conclude that the QD is not modified by the gate sensor bias point. This suggests that the poor phase response at large $V_\textup{GD}$ voltage cannot be explained by loss of confinement in the QD caused by an increased transparency of the barriers. By contrast, the origin of this effect should be attributed to an additional capacitive contribution arising from the formation of a 2DEG under GD, which dominates over the tunneling capacitance at high $V_\textup{GD}$. The data reported in Fig~\ref{cmp}(c) are consistent with this interpretation. Indeed, the signal excursion at $V_\textup{GD} =0.7$~V is more than an order of magnitude larger than the one at $V_\textup{GD} =0.5$~V, indicating a proportional variation of $\Delta C$ between the two operation modes. Furthermore, at lower $V_\textup{GD} $ the signal shows a nearly constant offset with sharp valleys corresponding to capacitance variations $\Delta C_\textup{t}\approx1.5$~aF due to tunnelling events. By contrast, at higher $V_\textup{GD}$ the signal varies steadily over the voltage range shown as an effect of an overall maximum variation in the total resonator capacitance  $\Delta C\approx60$~aF. The variation of the background signal is ascribable to the fact that, once $V_\textup{GD}$ is above threshold and a 2DEG is formed, the relevant MOS capacitance will depend on the voltages applied to nearby gates. In practice, the value of the additional capacitive contribution to the resonator is affected by each gate voltage in proportion to its electrostatic cross-coupling to the 2DEG. For example, in Fig.~\ref{cmp}(c) we show the cross-capacitive contribution of gate BL. Importantly, in this scenario, small capacitive contributions due to tunneling become exceedingly difficult to be detected since they are buried in this much larger capacitive swing, as highlighted in Fig.~\ref{cmp}(c).\\\indent
Next,  we characterize the change in total capacitance of the system as a function of $V_\textup{GD}$. We do so by monitoring the resonant frequency which directly relates to the total capacitance of the circuit as $f_\textup{R}=\frac{1}{2\pi \sqrt{LC}}$, where
 $C=C_\textup{p}+C_\textup{t}+C_\textup{MOS}$, and $C_\textup{MOS}$ is the conventional MOS gate capacitance for GD [see inset of Fig.~\ref{model}(a)]. By assuming that $C_\textup{p}$ should not change with gate voltage and neglecting the small contribution of $C_\textup{t}$, we extract the dependence of $C_\textup{MOS}$ with respect to $V_\textup{GD}$ from the resonant frequency shift. In Fig.~\ref{model}(a), we plot the extracted MOS capacitance as a function of $V_\textup{GD}$. For $V_\textup{GD}<0.6$~V we observe a small monotonic increase of the MOS capacitance as a function of increasing $V_\textup{GD}$. This phenomenon can be attributed to the charging of interface traps~\cite{koo} below the gate or to a cross-capacitive effect of gate GD to the neighbouring 2DEGs, such as those accumulated under SL and IG. For $V_\textup{GD}>0.6$~V the MOS capacitance rises rapidly, as expected from typical C-V curves of MOS capacitors in weak inversion~\cite{sze}. This type of behaviour is consistent with the accumulation of a 2DEG under the gate detector with a threshold voltage of $\approx0.6$~V.\\\indent
Interestingly, the data in Fig~\ref{model}(a) allow one to extract the change in capacitance caused by the rf signal for different $V_\textup{GD}$ values. The applied rf peak-to-peak voltage amplitude to the gate is $V_\textup{RF}=0.02$~V, hence we estimate  $\Delta C_\textup{MOS}\approx10$~aF at $V_\textup{GD}=0.7$~V. This confirms that modifications in the MOS capacitance of the gate sensor dominate at high values of $V_\textup{GD}$, making it increasingly difficult to detect the much smaller tunnelling capacitance variations due to single-electron transitions ($\Delta C_\textup{t}\approx1.5$~aF at the Coulomb peaks). By contrast, the good visibility of the tunnelling events at $V_\textup{GD}=0.5$~V is consistent with the fact that $\Delta C_\textup{MOS}$ is negligible at that operation point, as illustrated in Fig.~\ref{model}(a).
\\\indent In conclusion, we have demonstrated that, when an accumulation-mode gate sensor is operated above threshold voltage, the MOS capacitance becomes dominant with respect to the tunnelling capacitance, and charge readout degrades dramatically. 
In figures~\ref{model}(b) and (c), we suggest a way of circumventing this issue. By using a depletion gate (i.e. operated below threshold) which runs underneath all the dot gates, the accumulation of an electron layer is prevented except for the active region, where quantum dots are formed and controlled. Charge sensing can then be performed using the same gate that defines the QD via gate-based dispersive readout by embedding the gates into resonant circuits. Note that readout of every gate would not be strictly needed, as the state of each dot can be inferred via cross-correlated measurements of the nearest-neighbour sensors. To minimise the impact of the capacitance between dot and depletion gates on the dispersive response, we propose a design in which the depletion gate extends away from the active area for a distance $L$ until it reaches the transition region between gate (thin) and field (thick) SiO$_2$, of thickness $t$ and $h$, respectively. Note that in the field oxide region, the formation of a 2DEG underneath the dot gates is prevented by the larger oxide thickness ($h\gg t$) that results in a higher threshold voltage. We quantify the effect of the additional dot-to-depletion gate capacitance on the dispersive response by estimating the contribution to the overall parasitic capacitance for typical fabrication parameters. For dot gate widths $w=50$~nm and Al$_y$O$_x$ layer thickness of 2~nm, a phase response deterioration $\delta(\Delta\phi)/\Delta\phi$ of less than 1$\%$ can be attained for $L<$2~$\mu$m.
\\\indent This project has received funding from the European Union's Horizon 2020 research and innovation programme under grant agreement No 688539 and under the Marie Sk\l{}odowska-Curie grant agreement No 654712 (SINHOPSI). This work was also financially supported by the Australian National Fabrication Facility for device fabrication. ASD acknowledges support from the Australian Research Council (DP160104923 and CE11E0001017), the US Army Research Office (W911NF-13-1-0024) and the Commonwealth Bank of Australia. The authors acknowledge useful discussions with L. Hutin, D.A. Williams, A.J. Ferguson, F.E. Hudson and A.C. Betz.
\bibliography{ref_RF2DEG}

\end{document}